\documentclass[10pt, letter, twocolumn]{IEEEtran}

\usepackage{graphicx}
\usepackage{epsfig}
\usepackage{algorithm}
\usepackage{algorithmic}
\usepackage{ifmtarg}
\usepackage{caption}
\usepackage{footnote}
\usepackage{cite}
\usepackage{amssymb}
\usepackage{amsthm}
\usepackage[fleqn]{amsmath}
\usepackage{amsmath}
\usepackage{mathtools}
\usepackage{amsfonts}
\usepackage{epstopdf}
\usepackage{enumitem}
\usepackage{subfigure}
\usepackage{multirow}
\usepackage{color}
\usepackage[utf8]{inputenc}
\usepackage[english]{babel}

\usepackage[font=normalsize]{caption}
\theoremstyle{definition}

\usepackage[]{graphicx}
\usepackage{subfigmat}
\usepackage{changepage}
\def\BibTeX{{\rm B\kern-.05em{\sc i\kern-.025em b}\kern-.08em
    T\kern-.1667em\lower.7ex\hbox{E}\kern-.125emX}}

\usepackage[utf8]{inputenc}
\usepackage[english]{babel}

\begin{document}
\title{Energy Efficient D2D Communications Using Multiple UAV Relays}

\author{\vspace{-.4cm}
\IEEEauthorblockN{Ahmad Alsharoa$^{\text{1}}$ and Murat Yuksel$^{\text{2}}$}\\
\IEEEauthorblockA{$^{\text{1}}$Missouri University of Science and Technology (MST), Rolla, Missouri, USA. Email: aalsharoa@mst.edu}\\
\IEEEauthorblockA{$^{\text{2}}$University of Central Florida (UCF), Orlando, Florida, USA. Email:  murat.yuksel@ucf.edu}\vspace{-1.2cm}}

\maketitle
\thispagestyle{empty}
\pagestyle{empty}

\begin{abstract}
In this paper, we propose a novel optimization model for multiple Unmanned Areal Vehicles (UAVs) working as relays and helping Device-to-Device (D2D) communications at the same time. The goal of the UAVs is to operate in an energy-efficient manner while not only optimizing the available bandwidth and power allocations of the D2D links, but also act as relays when needed to maintain the communication between ground users. We formulate an optimization problem that maximizes the energy-efficient utility while respecting the resource availability including the UAVs' energy consumption, UAV-user association, and trajectory constraints. Due to the non-convexity of the problem, we propose to solve it in three steps using Taylor series approximation to optimize the power and the bandwidth, and use a heuristic algorithm for optimizing the UAVs' trajectory and UAV-user associations.
\end{abstract}

\maketitle

\vspace{-.4cm}
\section{Introduction}\label{Introduction}

Recently, Device-to-Device (D2D) communication has been considered an innovative feature of next-generation cellular networks.
It facilitates interoperability between close proximity wireless users using direct link and with minimal help from network infrastructures. More specifically, direct communication between nearby users enhances the spectrum utilization, overall throughput and data transfer rate, energy efficiency, and latency while enabling new peer-to-peer and location-based applications and services~\cite{D2D_su}. 

D2D communication becomes more challenging if a dynamic environment is considered.
When infrastructure support does not exist (e.g., during or after a disaster), LTE-Direct will not be available \cite{FCCreport}. UAVs can practically serve as base stations to organize and optimize communications among a swarm of devices while also acting as a relay to extend the devices' communication range.
In~\cite{dynamic}, a dynamic protocol was proposed to enable inter-cell D2D communication using a relay device.
The performance of the dynamic environment can be significantly improved by using UAVs to organize the D2D resources in the network and help in maintaining the communication links between out-of-range devices by working as a mobile relay.

As another dimension, optimizing the trajectories of UAVs supporting D2D communications among ground users can significantly enhance the network performance by determining the best coverage areas for the D2D links, and thus, optimize the resources.
Few works in the literature discuss the trajectory optimization of the UAVs in this context. For instance, in~\cite{UAVselfhealing}, Selim \textit{et. al} propose a novel trajectory optimization approach under a self-healing management framework, where multiple UAVs need to optimize their trajectories to heal the devices associated to a failed base station.
The UAV trajectory optimization using sequential convex optimization technique has been studied in~\cite{zhang1} for a point-to-point system model using only one UAV. In~\cite{zhang2}, the authors solve a one-dimensional placement problem and consider one UAV serving multiple ground users in a time sharing manner. This work simplifies the complexity of the optimization problem but limits applicability in practice. The work in~\cite{Hakim_UAV} proposed an energy efficiency management framework that optimizes the 3D trajectory of a UAV under cognitive radio system, where the authors proposed to optimize the 3D trajectory after minimizing the UAVs' energy consumption.

Coexistence between one UAV and underlaid D2D communication received notable attention. A downlink scenario is studied in~\cite{saad2}, where the UAV can serve only one device at a time and needs to travel from one stop point to another to serve other devices. The UAV does not manage or organize the D2D links, and considers it as interference to the UAV device. However, this approach limits the practicality of using a UAV as a base station and may consume a large amount of energy by forcing the UAV to travel from one point to another. In~\cite{Alsharoa_UAVDirect2}, we proposed for the first time an approach, called UAV-Direct, consisting of one UAV managing the resources of the D2D users.
However, this approach does not take the energy efficiency and UAVs' battery level into consideration. Furthermore, it is only limited to one UAV. The problem becomes more challenging when energy efficiency and multiple UAVs are considered serving multiple D2D users. Our proposed work in this paper considers multiple UAVs and their charging needs to serve D2D communications on the ground. This will firstly change the objective function and the goal of the problem and add more constraints in the resource allocation and trajectory optimization.

In this paper, we propose a new and novel energy-efficient framework where multiple UAVs optimize the power and bandwidth allocations for D2D users and work as relays when needed to maintain the communication links between users within or outside of communication range of each others. Thanks to their mobility, the UAVs are more robust against environmental changes and their trajectories can be optimized based on devices' dynamic locations.
To the best of our knowledge, energy-efficient optimization of trajectories for multiple UAVs that organize D2D communications is reported for the first time in this paper. 


\vspace{-.2cm}
\section{System Model}\label{SystemModel}

\begin{figure}[t!]
  \centerline{\includegraphics[width=2in]{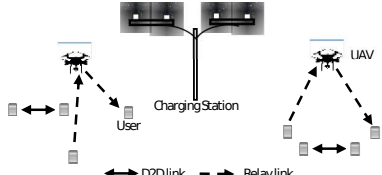}}
   \vspace{-3mm}
   \caption{System Model.}\label{system}
   \vspace{-3mm}
\end{figure}

We consider a wireless system composed with $L$ UAVs, mobile user pairs $u=1,..,U$ aiming to exchange data between each other using a direct D2D communication link (i.e., D2D users) or via one of the UAVs (i.e., relay users), and one UAV charging station located at the middle as shown in Fig.~\ref{system}.
We denote $N$ and $M$ as the total pairs of users that communicate using direct D2D links and UAV relay links, respectively.
We assume that the UAVs manage the resource allocations for both D2D and relay links.

We consider a 3D coordinate system where the coordinate of user $u$ and UAV $l$ at time instant $\zeta$ are given, respectively, as $\bold{W}_u(\zeta)= [x_u(\zeta), y_u(\zeta), 0]^t$ and $\bold{J}_l(\zeta)= [x_l(\zeta), y_l(\zeta), z_l(\zeta)]^t$, where $[.]^t$ denotes the transpose operator.
We assume that the time duration for the optimization $T$ is discretized into $t=1,..,\tilde{T}$ equal time slots such that $T=\tilde{T}\tau$, where $\tau$ is small enough that the movements of the users and the UAVs are negligible from our optimization problem's point of view. Note that the choice of $\tau$ depends on the mobility of the UAVs and users.
Without loss of generality, we assume that all UAVs cannot exceed their maximum speed denoted by $\bar{V}$. Therefore, the following trajectory constraints should be satisfied
\begin{equation}
||\bold{J}_l[t+1]-\bold{J}_l[t]||^2 \leq \bar{V} \tau, \forall l=1,.,L, t=1,.,\tilde{T},
\end{equation}
where $\bar{V} \tau$ is the maximum distances that the UAV $l$ can travel during each time slot $t$.
For simplicity, We assume that the total bandwidth $B$ is divided into two main fractions: $B^d$ for the D2D links and $B^r$ for the relay links. Further, we assume that all D2D links use $B^d$ at the same time while $B^r$ is divided to non-overlapping subfractions $B^r_m$ such that $\sum_{m=1}^{M} B^r_m=B^r$, where $B^r_m$ is the subfractional bandwidth assigned to the relay link $m$. This is considered a plausible assumption since the transmit power used of D2D users\footnote{We will sometimes refer to the set of direct D2D links in the system as `D2D users', which means the pairs of users communicating directly via D2D link.} is much less than the transmit power of the relay users\footnote{We will sometimes refer to the set of relay D2D links in the system as `relay users', which means the pairs of users communicating via a UAV.} due to the fact that the shadowing and fading for short range communications have much less effect on the channel gain compared to the large scale range of the relay links. Hence, using the same bandwidth for D2D users will not cause a large interference compared to relay users.
For relay user $m$, a binary variable $\epsilon^r_{lm}[t]$ is introduced, where it is equal to 1 if the relay user $m$ associated with UAV $l$ for the relay link at time $t$. 
We assume that each relay user can be associated
 with at most one UAV during time slot~$t$, on the other hand, each UAV can associate to multiple users. Thus, the following constraint should be respected: $\sum_{l=1}^L \epsilon^r_{lm}[t] \leq 1, \forall m, t.$

\vspace{-.5cm}
\subsection{Channel Model}
%

As discussed in~\cite{saad2} and~\cite{PL1}, the ground and air receivers can receive two types of signals in addition to the Line-of-Sight (LOS) signal. The first one is strong reflected Non Line-of-Sight (NLoS) signal and the second one is multiple reflected signal type that causes fading.
As shown in~\cite{PL1}, these types can be considered separately with different Probability of Occurrence (PoO).
In this case, there will be a probability to obtain a LoS link between the UAV and users in the relay link. The Path Loss (PL) between the UAV $l$ positioned at a position $\bold{J}_0$ and a ground user $m$ in urban environments for LoS and NLoS is given, respectively as~\cite{PL1}:
\small
\begin{equation}\label{LoS}
PL^{\text{LoS}}_{ml}[t]= \xi_{\text{LoS}} \left(\frac{4 \pi \delta_{ml}[t]}{\lambda_0}\right), PL^{\text{NLoS}}_{ml}[t]= \xi_{\text{NLoS}}\left(\frac{4 \pi \delta_{ml}[t]}{\lambda_0}\right),
\end{equation}
\normalsize
where $\delta_{ml}[t]=||\bold{J}_l-\bold{W}_m||$ is the distance between UAV $l$ and relay user $m$. $\lambda_0$ is the wavelength of the radio signal. $\xi_{\text{LoS}}$ and $\xi_{\text{NLoS}}$ are the additional loss to the free space propagation for LoS and NLoS links, respectively, due to the shadowing effect and the reflection of signals from obstacles.
The LoS probability is given by~\cite{PL2}, $p^{\text{LoS}}_{ml}[t]=1/(1+\nu_1 \exp(-\nu_2[\theta_{ml}[t]-\nu_1])),$
where $\theta_{ml}[t]=\frac{180}{\pi} \sin^{-1}\left(\frac{z_l[t]}{\delta_{ml}[t]}\right)$ is the elevation angle between the UAV and the user $m$ in degree. $\nu_1$ and $\nu_2$ are constant values that depend on the environment. The NLoS probability is, then, equal to $1-p^{\text{LoS}}_{ml}[t]$.
Using this PL model, the average PL for ground-to-air link is given by, $\text{PL}_{ml}[t]=p^{\text{LoS}}_{ml}[t] PL^{\text{LoS}}_{ml} [t]  +(1-p^{\text{LoS}}_{ml}[t]) PL^{\text{NLoS}}_{ml}[t].$
Therefore, the channel gain between user $m$ and the UAV in the relay link is given as $h^r_{ml}[t]=1/\text{PL}_{ml}[t].$
\vspace{-.5cm}
\subsection{UAV Power Model}
We consider both the transmission and operation power modes of the UAVs.
For the transmission power level, each UAV can be either in an active mode if it is in communication with one of the users or in an idle mode otherwise. For simplicity, the total transmit power consumption of UAV $l$ during a time slot $t$ to serve the associated relay users can be approximated by a linear model as~\cite{EARTH}, $P^r_l[t]=\alpha_l \sum_{m=1}^M \epsilon^r_{lm}[t] P^r_{l\acute{m}}[t]+\beta_l,$
where $\alpha_l$ corresponds to the power consumption that scales with the radiated power due to amplifier and feeder losses and $\beta_l$ models an offset of site power which is consumed independently of the average transmit power. $P^r_{l\acute{m}}[t]$ is the transmit power of UAV $l$ during time slot $t$ to forward the data from relay user $m$ to user $\acute{m}$.
Besides the power consumed for the transmission, the UAV consumes additional power for hovering and hardware, denoted by $P^f_l[t]$, and can be expressed as~\cite{Dpower_model} $P^f_l[t]= \sqrt{\frac{(m_\text{tot} g)^3}{2 \pi r_p^2 \omega_p \rho}}+\frac{P_\text{full}-P^s}{\bar{V}}V_l[t]+ P^s,$
where $m_\text{tot}$, $g$, and $\rho$ are the UAV mass in $\text{Kg}$, earth gravity in m$/\text{s}^2$, and air density in $\text{Kg}/\text{m}^3$, respectively. The parameter $r_p$ and $\omega_p$ are the radius and the number of the UAV's propellers, respectively.
$P_\text{full}$ and $P^s$ are extra hardware power consumptions when the UAV is moving at full speed and when it stops in a static position (i.e., $V_l[t] = 0$), respectively.
Thus, the total power consumption of UAV $l$ during time slot $t$ is given by $P_l[t]=P^r_l[t]+P^f_l[t].$

\vspace{-.5cm}
\subsection{Rate Calculation}
The transmission rate from relay user $m$ to the UAV in the relay link can be expressed as
\begin{equation}\label{Rate_m}
R^r_{ml}[t]=B^r_m[t] \log_2\left(1+\frac{P^r_{ml}[t] h^r_{ml}[t]}{B^r_m[t] N_0}\right),
\end{equation}
where $B^r_m[t]$ is the transmission bandwidth allocated to relay user $m$ in the relay link during time slot $t$, and $N_0$ is the noise power. $P^r_{ml}[t]$ is the transmit power of relay user $m$ to UAV $l$ during time slot $t$.
For simplicity and to make the problem more tractable, we assume that all relay users access the spectrum sparsely (allocate different bandwidth to different user, thus, no interference between users). 

Similarly, the transmission rate from  UAV $l$ to user $\acute{m}$ (the paired user) can be expressed as
\begin{equation}\label{Rate_mr}
R^r_{l\acute{m}}[t]=B^r_{m}[t] \log_2\left(1+\frac{P^r_{l\acute{m}}[t] h^r_{l\acute{m}}[t]}{B^r_{\acute{m}}[t]  N_0}\right).
\end{equation}
Therefore, the end-to-end maximum transmission rate at the destination (i.e., $\acute{m}$) using decode-and-forward (DF) approach where the UAVs decode the signals first before broadcasting it to the destination can be expressed as\cite{DFAlgamal} $R^r_m[t]= 1/2 \min\left(R^r_{ml}[t],R^r_{l\acute{m}}[t] \right).$

The transmission rate from user $n$ to the paired user $\acute{n}$ in the D2D link can be expressed as
\small
\begin{equation}\label{Rate_d}
R^d_n[t]=B^d \log_2\left(1+\frac{P^d_{n\acute{n}}[t] h^d_{n\acute{n}}[t]}{\sum_{\substack{k=1, k\neq n}}^N P^d_{k\acute{k}}[t] h^d_{k\acute{n}}[t]+ B^d N_0}\right),
\end{equation}
\normalsize
where $\sum_{\substack{k=1 \\ k\neq n}}^N P^d_{k\acute{k}}[t] h^d_{k\acute{n}}[t]$ is the interference power signal from other D2D users.

\section{Problem Formulation}\label{Formulation}
In this section, we formulate our optimization problems aiming to maximize the energy efficiency of the system respecting the UAVs' battery levels. In general, the total energy consumption of UAV $l$ during time slot $t$ can be expressed as
\begin{equation}\label{consumption}
E^c_l[t]=\tau \frac{||\bold{J}_l[t]-\bold{J}_0||^2}{||\bold{J}_l[t]-\bold{J}_0||^2+\tilde{J}} \left( P^f_l[t]  + P^r_l[t] \right)
\end{equation}
where $\bold{J}_0$ and $\tilde{J}$ are the location of the charging station and a very small number, respectively.
In~\eqref{consumption},  the ratio $||\bold{J}_l[t]-\bold{J}_0||^2/(||\bold{J}_l[t]-\bold{J}_0||^2+\tilde{J})$ is to ensure that the UAVs will not consume energy when located at the charging station. So it is 0 when UAV $l$ is in the charging station and very close to 1 otherwise.
On the other hand, we assume that UAV $l$ can be charged with a fix amount of power equal to $P^{ch}$ for each time instant when it is plugged to the charging station. Therefore, the total charging energy of UAV $l$ during time slot $t$ can be expressed as
\begin{equation}\label{charging}
E^{ch}_l[t]=\tau \left(1-\frac{||\bold{J}_l[t]-\bold{J}_0||^2}{||\bold{J}_l[t]-\bold{J}_0||^2+\tilde{J}}\right) P^{ch}.
\end{equation}
We assume that the UAVs are battery-powered devices. Therefore, the stored energy by UAV $l$ at the end of time slot $t$, denoted by $S_l[t]$, is given by $S_l[t]=S_l[t-1]+ E^{ch}_l[t]-E^c_l[t].$
We assume that, initially, each battery is charged by an amount of energy denoted by $S^0_l$.
In the sequel, we aim to maximize the energy efficiency utility of the system
by optimizing the followings parameters: 1) transmit power levels of the users and UAVs, 2) bandwidth allocation to each user, 3) association between UAVs and users, and 4) trajectory of the UAVs. Therefore, the optimization problem can be formulated as follows
\small
\begin{align}
&\hspace{-0.5cm}\underset{\substack{B^d[t], B^r_m[t], \bold{J}_l[t],\epsilon^r_{lm}[t]\\ P^r_{ml}[t], P^r_{l\acute{m}}[t], P^d_{n\acute{n}}[t] }}{\text{maximize}} \quad \mathcal{U}(R^d_{n}[t],R^d_m[t])\label{of}\\
&\hspace{-0.5cm}\text{subject to:}\nonumber\\
&\hspace{-0.5cm}  0 \leq P^d_{n\acute{n}}[t] \leq \bar{P}_u, \quad \forall n \quad \forall t,\label{powern}\\
&\hspace{-0.5cm}  0 \leq  P^r_{ml}[t] \leq \bar{P}_u, \quad  \forall m \quad \forall t,\label{powerm}\\
&\hspace{-0.5cm} \sum\limits_{m=1}^M \epsilon^r_{lm}[t] P^r_{l\acute{m}}[t] \leq \bar{P}_l,  \quad \forall l, \forall t,\label{power_r}
\end{align}
\begin{align}
&\hspace{-0.5cm}   B^d[t] + \sum\limits_{m=1}^M  \sum\limits_{l=1}^L \epsilon^r_{lm}[t] B^r_m[t] \leq \bar{B},  \quad \forall m, \forall t,\label{BW}\\
&\hspace{-0.5cm} \frac{||\bold{J}_l[t]-\bold{J}_l[0]||^2}{\bar{V}}P^f_l [t]|_{V_l=\bar{V}}+ E^c_l[t] \leq S_l[t-1], \quad \forall l, \forall t, \label{power_rbattery}\\
&\hspace{-0.5cm} S_l[t-1]+ E^{ch}_l[t] \leq \bar{S},  \quad \forall l, \forall t, \label{power_rbattery2}\\
&\hspace{-0.5cm}||\bold{J}_l[t+1]-\bold{J}_l[t]||^2 \leq V_l \tau, \quad \forall l, \forall n,\label{tra1}\\
&\hspace{-0.5cm} \sum\limits_{l=1}^L \epsilon^r_{lm}[t] \leq 1, \quad  \forall m, \forall t \label{Asso}
\end{align}
\normalsize
where $\mathcal{U}(R^d_n[t],R^d_m[t])$ denotes the energy efficiency utility of all users. Constraints~\eqref{powern}, \eqref{powerm},~and~\eqref{power_r} represent the peak power constraints at D2D users, relay users, and UAVs, respectively. Constraint~\eqref{BW} is to ensure the system bandwidth limitation.
Constraint~\eqref{power_rbattery} is also equivalent to $\sum_{\iota=1}^t E^c_l[\iota]- \sum_{\iota=1}^{t-1} E^{ch}_l[\iota] \leq S^0_l$, where the consumed energy is less than the stored energy in the previous time slot.
In constraint~\eqref{power_rbattery}, we assume that the UAVs' return speed to the charging station is $\bar{V}$. Thus, the term $\frac{||\bold{J}_l[t]-\bold{J}_l[0]||^2}{\bar{V}}P^f_l [t]|_{V_l=\bar{V}}$ is added to ensure that the UAV has enough battery to return to the charging station when needed. Constraint~\eqref{power_rbattery2} is equivalent to $S^0_l+ \sum_{\iota=1}^t E^{ch}_l[\iota] - \sum_{\iota=1}^{t-1} E^c_l[\iota] \leq \bar{S}$, where the charging energy that added to previous stored energy shouldn't exceed the UAV battery capacity (i.e., maximum stored energy $\bar{S}$).
Constraint~\eqref{tra1} indicates the trajectory constraint as explained in Section~\ref{SystemModel}.

In this work, we select to use Max-Min utility. The Max-Min utilities are a family of utility functions attempting to maximize the minimum data rate in the network~\cite{Min-Max}. By increasing the priority of users having lower rates, Max-Min utilities lead to more fairness in the network. In order to simplify the problem for this approach, we define a new decision variable $R_{\min}[t]=\underset{m,n}{\min}(R^d_n[t],R^d_m[t])$. Therefore, our optimization problem becomes:
\small
\begin{equation}\label{min_optimization}
\underset{\substack{B^d[t], B^r_m[t], \bold{J}_l[t],\epsilon^r_{lm}[t]\\ P^r_{ml}[t], P^r_{l\acute{m}}[t], P^d_{n\acute{n}}[t] }}{\text{maximize}} \quad   \frac{R_{\min}[t]}{ \sum\limits_{n=1}^N E^d_n[t] + \sum\limits_{m=1}^M E^r_{ml}[t]+  \sum\limits_{l=1}^L  E^c_l[t] } \vspace{-2cm}
\end{equation}
subject to:
\begin{align}
  &\hspace{-0.5cm} \frac{1}{2} R^r_{l\acute{m}}[t] \geq R_{\min}[t] \quad \forall m, \forall l, \forall t \label{min1}\\
 &\hspace{-0.5cm} \frac{1}{2} R^r_{ml}[t] \geq R_{\min}[t] \quad \forall m, \forall l, \forall t, \label{min2}\\
  &\hspace{-0.5cm} R^d_{n}[t] \geq R_{\min}[t] \quad \forall n, \forall t, \label{min3}\\
 &\hspace{-0.5cm} \eqref{powern}, \eqref{powerm},\eqref{power_r},\eqref{BW},\eqref{power_rbattery},\eqref{tra1},\eqref{Asso}, \label{other_constraint}
 \end{align}
 \normalsize
where $E^d_n[t]=\tau P^d_{n\acute{n}}[t]$ and $E^r_{ml}[t]= \tau P^r_{ml}[t]$.
\vspace{-.3cm}
\section{Proposed Solution}\label{Solution}
The formulated optimization problem is a non-convex problem due to constraints~\eqref{min_optimization}-\eqref{other_constraint}. We propose to solve it in three iterative steps.
At the beginning, we firstly optimize the power allocations by assuming fixed bandwidths, associations, and UAV trajectories. In this step, we approximate the solution by converting our formulated problem to a convex one. We secondly optimize the bandwidth allocations for both D2D and relays users with a similar approximation technique. Finally, we employ a recursive heuristic search algorithm to optimize the UAV trajectories and UAV-user associations together. These steps are repeated until convergence.
\vspace{-.3cm}
\subsection{Transmit Power Allocations}
For fixed bandwidth allocations and UAV trajectories, the optimization problem can be given as
%
\begin{align}\label{min_optimization2}
&\hspace{-0.5cm}\textbf{(\text{P1}):} \underset{\substack{P^r_{l,m}[t],P^r_{l\acute{m}}[t], \\P^d_{n\acute{n}}[t],R_{\min}[t] }} {\text{maximize}} \quad \mathcal{U}(R_{\min}[t])
&\hspace{-0.5cm}\text{subject to:} \eqref{min1}-\eqref{other_constraint}.
\end{align}

\noindent This problem is quasi-concave (i.e., 1/(objective function) is quasi-convex) except constraint~\eqref{min3} since its objective function, $\mathcal{U}(R_{\min}[t])$, is a fraction of a concave function and a linear function and the other constraints are convex.
Hence, the goal is to convert constraint \eqref{min3} into a convex one in order to solve the problem efficiently.
This constraint is neither concave nor convex with respect to $P^d_{n\acute{n}}$. We can expand the left hand side of constraint~\eqref{min3} as follows:
\small
\begin{align}\label{power_non0}
R^d_n[t]
 =&\underbrace{B^d[t] \log_2\left(B^d[t] N_0+\sum\limits_{k=1}^N P^d_{k\acute{k}}[t] h^d_{k\acute{n}}[t]\right)}_{\tilde{R}^d_{n,1}[t]}
\\&
\underbrace{-B^d[t] \log_2\left( \sum\limits_{\substack{k=1 \\ k\neq n}}^N  P^d_{k\acute{k}}[t] h^d_{k\acute{n}}[t]+ B^d[t] N_0\right).}_{\tilde{R}^d_{n,2}[t]}
   \nonumber
\end{align}
\normalsize
Now, the main goal is to convert \eqref{power_non0} to a concave form in order for \textbf{P1} to become convex.
Note that $\tilde{R}^d_{n,1}[t]$ is concave, because the $\log$ of an affine function is concave~\cite{Boyd}. Also, $\tilde{R}^d_{n,2}[t]$ is a convex function, and thus, it needs to be converted to a concave function. To tackle the non-concavity of $\tilde{R}^d_{n,2}[t]$, the Successive Convex Approximation (SCA) technique can
be applied where in each iteration, the original function is approximated by a more tractable function at a given local point. Recall that $\tilde{R}^d_{n,2}[t]$ is convex in $P^d_{n\acute{n}}[t]$, and any convex function can be globally lower-bounded by its first order Taylor expansion at any point. Therefore, given $P^d_{n\acute{n}}(r)[t]$ in iteration $r$, we obtain the following lower bound for $\tilde{R}^d_{n,2}(r)[t]$:
\begin{align}\label{power_con}
\tilde{R}^d_{n,2}(r)[t] \geq & -B^d[t] \log_2\left(\psi(r)[t] \right) \nonumber \\& -\frac{h^d_{k\acute{n}}[t]}{\ln(2)\psi(r)[t]}(P^d_{k\acute{k}}[t] - P^d_{k\acute{k}}(r)[t])
\end{align}
where $\psi(r)[t]=\sum_{\substack{k=1, k\neq n}}^N  P^d_{k\acute{k}}(r)[t] h^d_{k\acute{n}}[t]+ B^d[t] N_0$.

At this stage, \textbf{P1} is a quasi-convex optimization. Hence, its solution is equivalent to finding
the root of the scalar function $\mathcal{U}(R_{\min}[t])=1/\mathcal{F}(\kappa)$, which can be solved using the bisection method. Here, $\mathcal{F}(\kappa)$ is a convex, continuous, and strictly decreasing function with respect to $\kappa$,
and is defined as $\mathcal{F}(\kappa)=\min \left(\mathcal{N}-\kappa \mathcal{D} \right)$,
where $\mathcal{N}$ and $\mathcal{D}$ represent the nominator and denominator of $\mathcal{F}(\kappa)$, respectively.
The last step to solve \textbf{P1} is to apply SCA to find the best approximation of constraint~\eqref{power_con}.

\vspace{-.3cm}
\subsection{Bandwidth Allocations}\label{app_BW}
For given power allocations, association, and UAV trajectories, the problem for optimizing the bandwidth allocations can be given as
\begin{equation}\label{min_optimization3}
   \textbf{(\text{P2}):} \underset{\substack{B^d[t], B^r_m[t],R_{\min}[t] }} {\text{maximize}} \quad \mathcal{U}(R_{\min}[t])
\vspace{-3mm}
\end{equation}
\hspace{0.5cm} subject to: \eqref{min1}-\eqref{other_constraint}.

\noindent The objective function~\eqref{min_optimization3} is quasi-convex and all constraints of \textbf{P2} are convex functions except~\eqref{min1}-\eqref{min3}.
These constraints are neither concave nor convex with respect to the bandwidth allocations.
In the sequel, we approximate constraint~\eqref{min3} to a convex function. The same approximation approach can be applied for the other two constraints, i.e.,~\eqref{min1} and~\eqref{min2}. \footnote{We omit the details of these approximations due to space limitations.}
The left hand side of~\eqref{min3} can be expanded as follows
\small
\begin{align}\label{power_non}
R^d_n[t]
 =&\underbrace{B^d[t] \log_2\left(B^d[t] N_0+\sum\limits_{k=1}^N P^d_{k\acute{k}}[t] h^d_{k\acute{n}}[t] \right)}_{\tilde{R}^d_{n,1}[t]}
\\&
\underbrace{-B^d[t] \log_2\left( \sum\limits_{\substack{k=1 \\ k\neq n}}^N  P^d_{k\acute{k}}[t] h^d_{k\acute{n}}[t]+ B^d[t] N_0\right).}_{\tilde{R}^d_{n,2}[t]}
   \nonumber
\end{align}
\normalsize

To prove the quasi-convexity of the optimization problem formulated in \textbf{P2}, we need to prove that both $\tilde{R}^d_{n,1}[t]$ and $\tilde{R}^d_{n,2}[t]$ are concave.
Let us start with $\tilde{R}^d_{n,2}[t]$. We refer to the following lemma in~\cite{Boyd}:

\noindent \textbf{Lemma 1}: If $f$ and $g$ are concave, positive, with one non-decreasing and the other non-increasing, then $fg$ is concave.

It can be noticed that $-B^d[t]$ is concave and non-increasing while $\log_2\left( \sum_{\substack{k=1, k\neq n}}^N  P^d_{k\acute{k}}[t] h^d_{k\acute{n}}[t]+ B^d[t] N_0\right)$ is concave and non-decreasing in terms of $B^d[t]$. Hence, $\tilde{R}^d_{n,2}[t]$ is a concave function.

Using the same approach, we can prove that $\tilde{R}^d_{n,1}[t]$ is a convex function (i.e., by proving that $-\tilde{R}^d_{n,1}[t]$ is a concave function). Therefore, $\tilde{R}^d_{n,1}[t]$ needs to be converted to a concave function in order to make constraint~\eqref{min3} concave.
To tackle the non-concavity of $\tilde{R}^d_{n,1}[t]$, the SCA technique can
be applied (similar to \textbf{P1}) where in each iteration, the original function is approximated by its first order Taylor expansion. Therefore, given $B^d(r)[t]$ in iteration $r$, we obtain the following lower bound for $\tilde{R}^d_{n,1}$(r)[t]:
\begin{align}\label{B_con}
\hspace{-5mm}\tilde{R}^d_{n,1}(r)[t] \geq & \phi_2(r)[t]+ \\& \left[\frac{B^d(r)[t] N_0}{\ln2\phi_1(r)[t]} + \phi_2(r)[t] \right] (B^d[t]-B^d(r)[t])\nonumber.
\vspace{-10mm}
\end{align}
\normalsize
where $\phi_1(r)[t]=B^d(r)[t] N_0+\sum_{k=1}^N P^d_{k\acute{k}}[t] h^d_{k\acute{n}}[t]$ and $\phi_2(r)=B^d(r) \log_2(\phi_1(r))$.
By applying the same procedure to the other constraints (i.e., \eqref{min1} and \eqref{min2}), \textbf{P2} becomes a quasi-convex optimization and it can be solved efficiently using SCA.


\vspace{-.3cm}
\subsection{UAV Trajectories and Association Optimization}\label{algo}
In this subsection, we consider optimizing the trajectories and associations of the UAVs for fixed resource allocations (i.e., transmit powers and bandwidth allocations).
Even with fixed resource allocations, the problem is still non-convex and it is very difficult to find an approximate solution due to the channel expression and the association binary variables $\epsilon^r_{lm}[t]$. Therefore, we introduce a quick and efficient algorithm based on a recursive shrink-and-realign process. The main advantages of this algorithm over other heuristic algorithms can be summarized as follows: (i) it is easy to implement by using a simple search process with few parameters to manipulate, (ii) it has low computational cost, and (iii) it provides fast convergence to a close-to-optimal solution.


We propose a Recursive Uniform Search (RUS) algorithm to optimize the UAV trajectories and associations. We assume that the association between UAVs and users can be done based on the best favorite channel, where the user is associated with the best UAV link.
Our algorithm starts by generating initial $Q$ high-efficiency next position candidates for each UAV $\bold{J}^{q}_l,\;q=1 \cdots Q, \forall l$ with a total of $Q^L$ candidates for all UAVs to identify promising candidates and to form initial populations ${\mathcal Q}$. We select to distribute the candidates uniformly over the surface of a sphere (we start by assuming the radius of this sphere $r_0$ equal to half of the UAV coverage radius) and the initial candidate is its center (where the UAV is currently located). Then, it determines the objective function achieved by each candidate by solving \textbf{P1} and \textbf{P2}, and this will guide us to the direction of the best candidate. Note that the association depends on the trajectory of each UAV and also depends on the channel between UAVs and users.
After that, it finds the initial best local candidate $q^{i,\text{local}}[t]$ that provides the best solution for iteration $i$.
Then, we start recursive sampling with uniform distribution over a new sphere with a radius \emph{shrunk} to half of the previous sphere (i.e., $r_{i-1}/2$) and its center realigned to $q^{i,\text{local}}[t]$. Using this shrink-and-realign process, we find the best solution $q^*$ and the corresponding trajectory $\bold{J}^{q^*}$. This shrink-and-realign procedure is repeated until the size of the sample space, i.e., the volume of the sphere, decreases below a certain threshold or reach a maximum iteration count of $I_\text{iter}$.
The details of the joint optimization approach are given in Algorithm~\ref{joint}.

\begin{algorithm}[h!]
\caption{Joint optimization algorithm}\label{joint}
\small
\begin{algorithmic}[1]
\STATE i=1.
\STATE Generate an initial population ${\mathcal Q}$. 
\WHILE {{Not} converged or reaching maximum iteration}
\FOR {$q=1 \cdots Q^L$}
\STATE Initialize $P^r_{ml}[t], P^r_{l\acute{m}}[t], P^d_{n\acute{n}}[t], B^d[t], B^r_m[t]$
\WHILE {{Not} converged}
\STATE Find $P^r_{ml}[t], P^r_{l\acute{m}}[t], P^d_{n\acute{n}}[t], B^d[t], B^r_m[t]$ by solving \textbf{P1} and \textbf{P2} optimization problems for candidate $q$ after using the approximation approaches described in Sections~\ref{Solution}.
\ENDWHILE
\ENDFOR
\STATE Find $q^{\text{i,local}}[t]=\arg\underset{q}{\mathrm{max}}\, \mathcal{U}(R_{\min}[t])$, (i.e., $q^{\text{i,local}}[t]$ indicates the index of the best local candidate that results in the highest objective function for iteration $i$).
\STATE Initially $r_0=\bar{V} \tau$.
\IF {$q^{\text{i,local}}[t] \leq q^{\text{i-1,local}}[t]$}
\STATE  Re-align the center of sample space to the new point.
\STATE Start recursive sampling with uniform distribution over a sphere with center equal to $q^{i,\text{local}}[t]$.
\ELSE
\STATE Shrink the sample space by updating the radius.
\ENDIF
\STATE i=i+1.
\ENDWHILE
\end{algorithmic}
\normalsize
\end{algorithm}

\vspace{-.4cm}
\section{Selected Numerical Results}\label{Simulations}
In this section, we provide selected numerical results to show our system's performance. We consider a system with $L=5$ UAVs flying at 60 m elevation, connected with different number of ground users distributed randomly within an area of 800m $\times$ 800m.
We assume that $\bar{P}_l$ is the same for all UAVs and $\bar{P}_u$ is the same for all ground users.
The noise power $N_0$ is assumed to be $2.5\times 10^{-25}$ W/Hz~\cite{saad2}. The constant values are selected to be $\nu_1=9.6$ and $\nu_2=0.29$ for a low-elevation atmosphere where the UAVs will be flying~\cite{saad2}.

The charging station is located at the center (400m, 400m, 60m).
Initially we assume that the locations, i.e., x and y coordinates, of the UAVs are in meters as [(400,400,60), (200,200,60), (200,600,60), (600,200,60), (600,600,60)]. Also, we assume that the initial battery level of the UAVs are given respectively as ($\bar{S},\bar{S}/2,\bar{S}/2,\bar{S}/2,\bar{S}/2$).
In Table~\ref{tab2}, we present the values of the remaining environmental parameters used in the simulations, which are found to be representative for low-flying UAVs~\cite{Alsharoa_drone}.
{\small
\begin{table}[h!]
\centering
\caption{\label{tab2} System parameters}
\vspace{-1mm}
\addtolength{\tabcolsep}{-2pt}\begin{tabular}{|l|c||l|c||l|c|}
\hline
\textbf{Parameter} & \textbf{Value} & \textbf{Parameter} & \textbf{Value}& \textbf{Parameter} & \textbf{Value}\\ \hline \hline
 $\bar{B}$(MHz)  & $20$ & $\xi_{\text{LoS}}$ (dB) & 1 & $\xi_{\text{NLoS}}$ (dB) & 12 \\ \hline
 $\lambda$ (m) & 0.125 & $Q$  & $10$ & $I_\text{iter}$ & 10 \\ \hline
$\alpha_l$ & 4 & $\beta_l$ (W) & 6.8 & $\bar{S}$ (kJ) & $15$ \\ \hline
$r_p$ (cm) & $20$ & $\omega_p$ & 4 & $P^s$ (W) & 0.5\\ \hline
$P^{ch}$ (W) & $10$ & $\bar{V}$ (m/s) & 15 & $m_{\text{tot}}$ (Kg) & 1 \\ \hline
\end{tabular}
\vspace{-1mm}
\end{table}}

\begin{figure}[h!]
  \centerline{\includegraphics[width=3.5in]{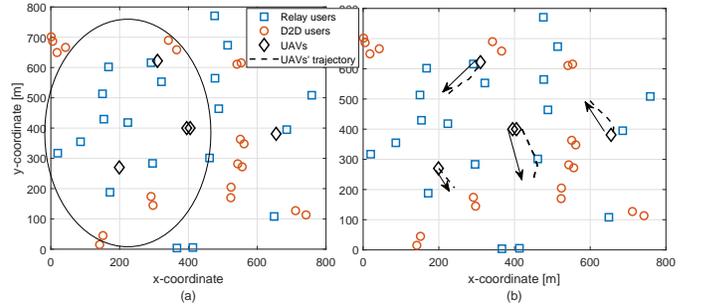}}
   \caption{The proposed trajectory example with $\bar{P}_l=36$ dBm, $\bar{P}_u=20$ dBm, $M=10$ and $N=10$.}\label{Trajectory}
   \vspace{-5mm}
\end{figure}

\begin{figure}[h!]
  \centerline{\includegraphics[width=3.5in]{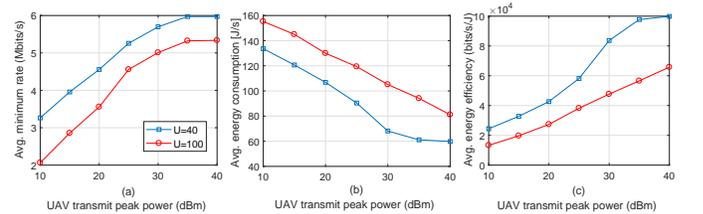}}
   \caption{System performance versus UAV transmit power $\bar{P}_l$ with fixed $\bar{P}_u=20$ dBm, for two different user densities.
}\label{throughput}
   \vspace{-5mm}
\end{figure}

%

We start our simulation results by showing how the UAV trajectories update if users move to new locations. Fig.~\ref{Trajectory} considers an example of this effect. It shows that when some users move to certain directions, the UAVs update their trajectories to maintain an energy-efficient solution.
We assumed that for a given time, the locations of users are given in Fig.~\ref{Trajectory}-a. Based on the optimization solution, only 3 out of 5 UAVs are needed to maximize the energy efficiency for these user locations. However, when users start moving, such as some of them moving to the left side (i.e., the users inside the oval shape in Fig.~\ref{Trajectory}-b), we update the optimization problem based on the battery level of the UAVs. This may increase the number of needed UAVs as shown in Fig.~\ref{Trajectory}-b (in this case a fourth UAV is moving from the charging station in the center to an optimize serving location.)
This proves that our solution is a dynamic solution taking into consideration the users' movements.

The system performance is discussed in Fig.~\ref{throughput}. 
We consider low and high user density scenarios, $U$=40 and $U$=100, respectively, where we assume $M=N$ in both scenarios and users' power budget is $\bar{P}_u=20$ dBm.
In Fig.~\ref{throughput}-a, we plot the average minimum user throughput versus UAVs' transmit power $\bar{P}_l$. In terms of minimum throughput, we notice that the minimum achievable throughput is improving with the increase of $\bar{P}_l$ up to a certain point and then it remains almost constant, due to the fact that the minimum throughput also depends on the $R^r_{m,l}$ (the uplink between the user $m$ and UAV $l$) that is dependent on $P^r_{m,l}$ and limited by $\bar{P}_u$.
On the other hand, Fig.~\ref{throughput}-b plots the average energy J/s versus $\bar{P}_l$. We have two remarks, firstly, serving more users increases the energy consumption, and secondly, as $\bar{P}_l$ increases, the coverage area of the UAVs increases which allow more flexibility in sending UAVs from charging station to serve users.  Fig.~\ref{throughput}-c plots the average energy efficiency versus $\bar{P}_l$. 

\begin{figure}
\begin{minipage}[c]{0.48\linewidth}
\includegraphics[width=1.85in]{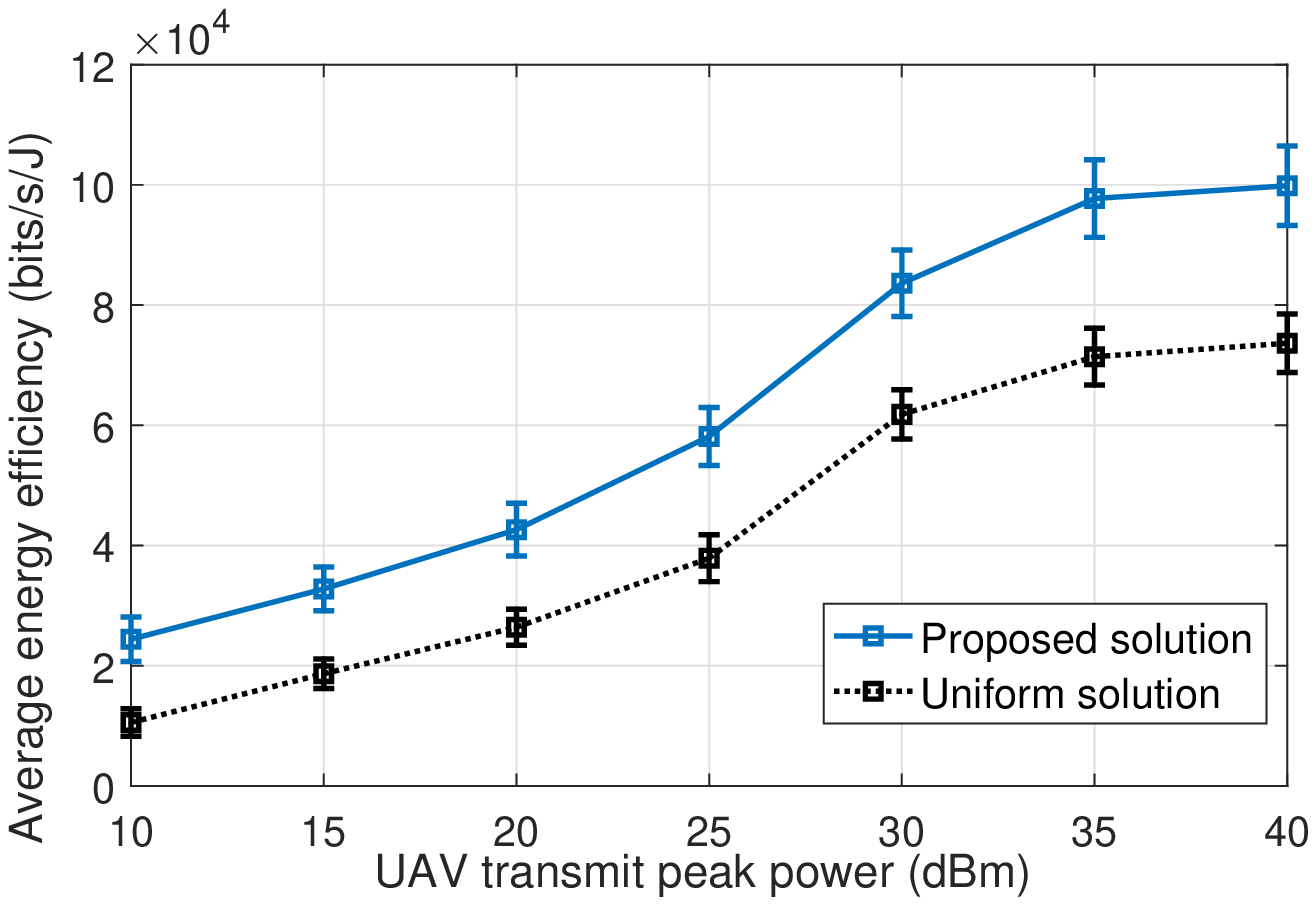}
\caption{Proposed versus uniform solutions.}\label{EE_updated2}
\end{minipage}
\hfill
\begin{minipage}[c]{0.48\linewidth}
\vspace{-.2cm}
\includegraphics[width=1.7in]{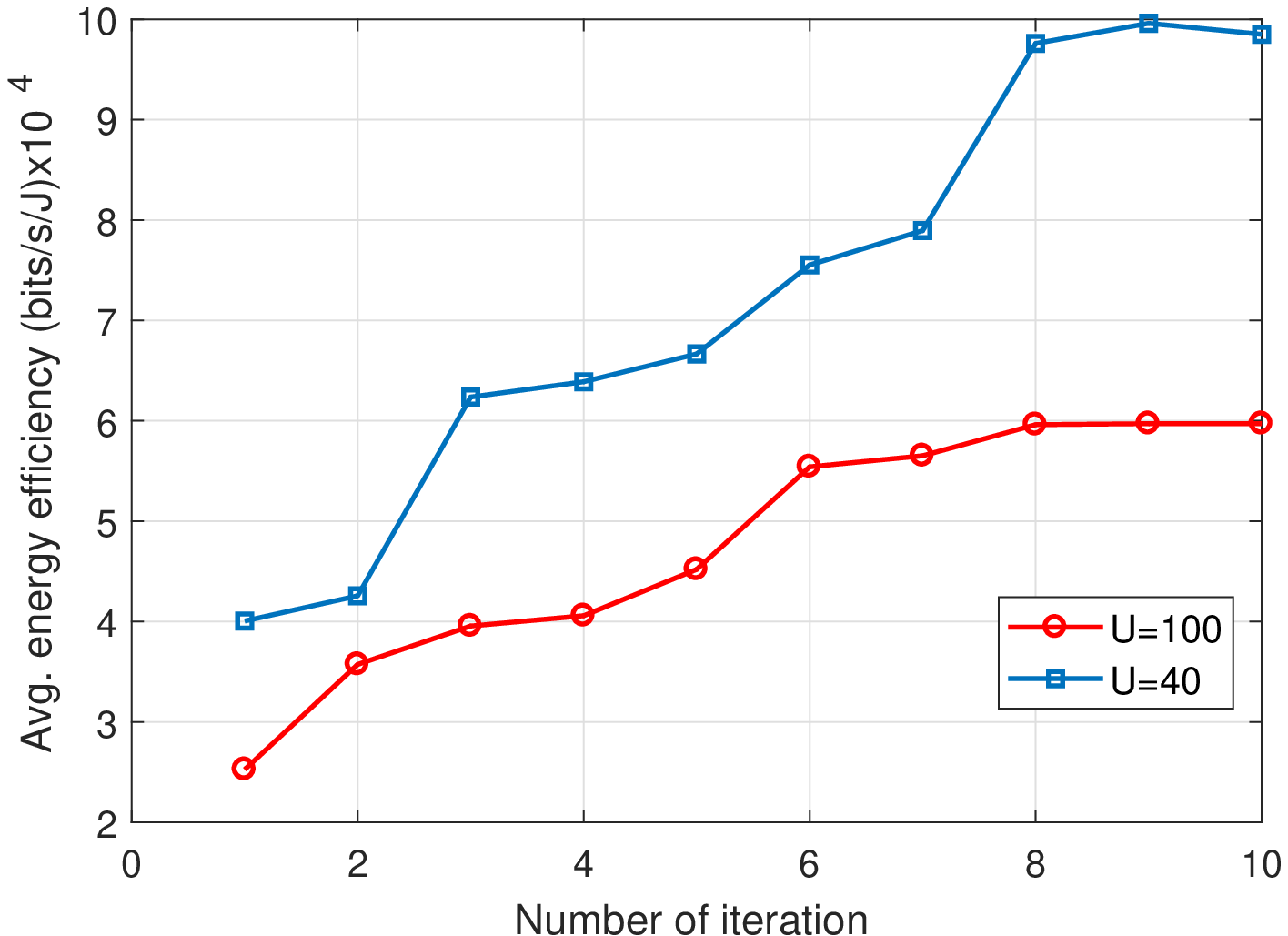}
\caption{Convergence speed.}\label{convergence}
\end{minipage}%
\vspace{-8mm}
\end{figure}

Fig.~\ref{EE_updated2} compares between our proposed solution and uniform case, where we distribute the power and bandwidth equally to the users. Please note that, we assume fixed power $\bar{P}_u$ for simplicity only. The figure shows that our proposed solution outperforms the uniform case. For instance, our proposed solution can achieve 50\% improvement in energy efficacy compared to uniform case by using $\bar{P}_l=25$ dBm by acheiving around $6\times10^4$ bits/s/J instead of $4\times10^4$ bits/s/J. In addition, we can see that the confidence interval is around 94\%.


Finally, the convergence speed of our approximation is shown in Fig.~\ref{convergence}. We plot the energy efficiency versus the number of iterations for two user densities to solve the optimization problems \textbf{\text{P1}} and \textbf{\text{P2}} for $\bar{P}_l=36$ dBm, and $\bar{P}_u=20$ dBm. The results shows that we can reach convergence with only few iterations (fewer than 8 iterations). Note that an iteration in Fig.~\ref{convergence} corresponds to one iteration of the ``while loop'' given in
Algorithm 1 (i.e., line 6-8). This implies that the UAVs will be able to calculate their near-optimal resource allocation in real-time as each iteration of the while loop takes a small amount of time, i.e., 1-2 seconds with a typical CPU.

\section{Conclusion}\label{Conclusions}
In this paper, we proposed a new approach to manage the resource allocation for D2D communications using multiple moving UAVs. The UAVs can work as relays when needed.
We formulated an optimization problem that maximizes the energy efficient utility while taking into consideration the power and bandwidth limitations, in addition to the association constraints.
The optimization framework enables the UAVs to optimize their trajectories as well as the transmit power and bandwidth allocations of the D2D links while also deciding which user devices on the ground are going to be associated to which UAV.
Due to non-convexity of the problem, we proposed an approximated solution based on Taylor series expansion for resource allocation and a recursive shrink-and-realign process for trajectory and UAV-user association optimization. In our next challenging task,
we are going to improve our system model by considering multi-hop relays among the multiple UAVs trying to facilitate D2D communications among the user devices on the ground. This will add more complexity to the problem, but on the other hand, it will further improve the performance.

\bibliographystyle{IEEEtran}
\bibliography{C_2020ICC}

\end{document}